\documentclass[aps,prl,twocolumn,showpacs,superscriptaddress]{revtex4}
\usepackage{graphicx}

\begin{document}

\title{Strong-Coupling Superconductivity in NaFe$_{1-x}$Co$_x$As: the Eliashberg Theory and Beyond}

\author{Guotai Tan}
\affiliation{Beijing Normal University, Beijing 100875, China}
\affiliation{Department of Physics and Astronomy, The University of Tennessee, Knoxville, Tennessee 37996-1200, USA}
\author{Ping Zheng}
\affiliation{Beijing National Laboratory for Condensed Matter Physics, Institute of Physics, Chinese Academy of Sciences, Beijing 100190, China}
\author{Xiancheng Wang}
\affiliation{Beijing National Laboratory for Condensed Matter Physics, Institute of Physics, Chinese Academy of Sciences, Beijing 100190, China}
\author{Yanchao Chen}
\affiliation{Beijing National Laboratory for Condensed Matter Physics, Institute of Physics, Chinese Academy of Sciences, Beijing 100190, China}
\author{Xiaotian Zhang}
\affiliation{Beijing National Laboratory for Condensed Matter Physics, Institute of Physics, Chinese Academy of Sciences, Beijing 100190, China}
\author{Jianlin Luo}
\affiliation{Beijing National Laboratory for Condensed Matter Physics, Institute of Physics, Chinese Academy of Sciences, Beijing 100190, China}
\author{Tucker Netherton}
\affiliation{Department of Physics and Astronomy, The University of Tennessee, Knoxville, Tennessee 37996-1200, USA}
\author{Yu Song}
\affiliation{Department of Physics and Astronomy, The University of Tennessee, Knoxville, Tennessee 37996-1200, USA}
\author{Pengcheng Dai}
\affiliation{Department of Physics and Astronomy, The University of Tennessee, Knoxville, Tennessee 37996-1200, USA}
\affiliation{Beijing National Laboratory for Condensed Matter Physics, Institute of Physics, Chinese Academy of Sciences, Beijing 100190, China}
\author{Chenglin Zhang}
\affiliation{Department of Physics and Astronomy, The University of Tennessee, Knoxville, Tennessee 37996-1200, USA}
\author{Shiliang Li}
\email{slli@iphy.ac.cn}
\affiliation{Beijing National Laboratory for Condensed Matter Physics, Institute of Physics, Chinese Academy of Sciences, Beijing 100190, China}
\begin{abstract}
We study the normal-state and superconducting properties of NaFe$_{1-x}$Co$_x$As system by specific heat measurements. Both the normal-state Sommerfeld coefficient and superconducting condensation energy are strongly suppressed in the underdoped and heavily overdoped samples. The low-temperature electronic specific heat can be well fitted by either an one-gap or a two-gap BCS-type function for all the superconducting samples. The ratio $\gamma_NT_c^2/H_c^2(0)$ can nicely associate the neutron spin resonance as the bosons in the standard Eliashberg model. However, the value of $\Delta C/T_c\gamma_N$ near optimal doping is larger than the maximum value the model can obtain. Our results suggest that the high-$T_c$ superconductivity in the Fe-based superconductors may be understood within the framework of boson-exchange mechanism but significant modification may be needed to account for the finite-temperature properties.
\end{abstract}


\pacs{74.70.Dd}

\maketitle

Strong-coupling superconductivity in the conventional superconductors can be well described within the framework of Eliashberg theory \cite{CarbotteJP90}, where electron Cooper pairs are mediated by virtual phonons or some other 
bosons. Since spin fluctuations may act as the mediating bosons for electron pairing and superconductivity \cite{EschrigM06,scalapino},
it is important to determine if the Eliashberg-based theory can understand the transport and magnetic properties of unconventional superconductors \cite{CarbotteJP11}. 
For copper oxides, this is difficult due to the plethora of phases competing with superconductivity
 and the $d$-wave nature of the superconducting gap symmetry.
The Fe-based superconductors may offer a better opportunity to test the suitability of 
the Eliashberg theory due to the $s$-wave nature of the superconducting electron pairing 
and the Fermi-liquid-like normal states \cite{hirschfeld}.  

In the standard Eliashberg theory, the superconducting electron Cooper pairs are mediated by bosons with an average energy of $\omega_{ln}$. For a $\delta$-function electron-boson spectral density $\alpha^2F(\omega)$ = $A\delta(\omega-\omega_E)$, we have $\omega_{ln}$ = $\omega_E$.  The ratio of $T_c/\omega_{ln}$ representing the coupling strength 
is related to two important dimensionless parameters $\gamma_NT_c^2/H_c^2(0)$ and $\Delta C/T_c\gamma_N$, where $\gamma_N$, $H_c(0)$ and $\Delta C/T_c$ are the normal-state Sommerfeld coefficient, the thermodynamic critical field at zero temperature, 
and the specific heat jump across $T_c$, respectively \cite{CarbotteJP90}. 
For conventional superconductors, these two ratios can be solved analytically through 
\begin{eqnarray}
\frac{\gamma_NT_c^2}{H_c^2(0)} &= 0.168\left[1-12.2\left(\frac{T_c}{\omega_{ln}}\right)^2ln\left(\frac{\omega_{ln}}{3T_c}\right)\right],\\
\frac{\Delta C}{T_c\gamma_N} &= 1.43\left[1+53\left(\frac{T_c}{\omega_{ln}}\right)^2ln\left(\frac{\omega_{ln}}{3T_c}\right)\right].
\end{eqnarray} 
\noindent We see that these two ratios has a linear relationship and 
should be simultaneously satisfied for a given superconductor.

Recently, the bosonic spectrum is found in the tunneling measurements on the Fe-based superconductors \cite{FasanoY10,ShanL12,WangZ12}. It is thus important to 
determine to what extent the standard Eliashberg theory holds by checking the validity of Eqs. (1) and (2). 
The thermodynamic properties of the Fe-based superconductors have been measured in many systems \cite{LiZ08,BudkoSL09,MuG09,PopovichP10,GofrykK10a,HardyF10,WeiF10,KimJS11,StewartGR11,HuJ11,NojiT12,WangAF12}. 
In Ba$_{0.6}$K$_{0.4}$Fe$_2$As$_2$, a calculation based on the Eliashberg model considering multiple bands is able to quantitatively reproduce the experimental results 
based on the assumption that spin fluctuations are electron pairing mediating bosons
\cite{PopovichP10}. Recently, a very sharp neutron spin resonance is found in superconducting NaFe$_{0.955}$Co$_{0.045}$As \cite{ZhangC12}.
The mode, centering at the in-plane antiferromagnetic wave vector, is strictly two-dimensional in the reciprocal space, which leads to an easy way of considering $\alpha^2F(\omega)$ and hence $\omega_{ln}$. Therefore, the NaFe$_{1-x}$Co$_x$As system may be suitable to check the Eliashberg theory. 

In this paper, we report a comprehensive study on the electron-doping evolution of the specific heat in NaFe$_{1-x}$Co$_x$As. 
The measured value of $\gamma_NT_c^2/H_c^2(0)$ is consistent with that obtained from Eq. (1) by assuming that $\omega_{ln}$ is equal to the neutron spin resonance energy \cite{ZhangC12}. However, the value of $\Delta C/T_c\gamma_N$ reaches up to 3.7 near optimal doping, which is much larger than the maximum value of Eq. (2). Our results suggest that the high-$T_c$ superconductivity in the Fe-based superconductors may be understood within the conventional boson-exchange mechanism but the finite-temperature properties should be revised around the optimal doping.

Single crystals of NaFe$_{1-x}$Co$_x$As were grown by the self-flux method as reported previously \cite{TanatarMA12}. The samples were attached onto the heat capacity pucks in the glovebox and transported within a sealed bottle to avoid the sample quality change \cite{TanatarMA12}. 
The time that the samples were exposed to air during the installation of the puck was less than 1 minute. The specific heat was measured by the PPMS from Quantum Design. 

\begin{figure}[tbp]
\includegraphics[scale=.43]{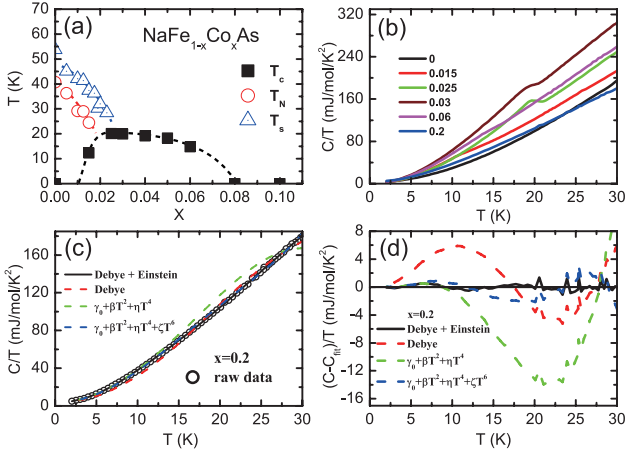}
\caption{(Color online) (a) Phase diagram of NaFe$_{1-x}$Co$_x$As that shows $T_c$ (black solid square), $T_N$ (red open circle) and $T_s$ (blue open triangle). The dash lines are guided to the eye. (b) Specific heats of some samples plotted as C/T vs T. (c) Fitted results of several models on the specific heat of the x = 0.2 sample. The differences between the data and each model are given in (d). 
}
\label{fig1}
\end{figure}

The phase diagram of NaFe$_{1-x}$Co$_x$As is very similar to other iron pnictides 
with a long-range AF order in the parent compound and a dome-like superconducting regime \cite{dai,ParkerDR10,WangAF12,LiS09}, as shown in Fig. \ref{fig1}(a). The structural transition temperature $T_s$ and magnetic transition temperature $T_N$ determined from the resistivity measurement are similar to those reported in the other literatures \cite{ParkerDR10,WangAF12}. The $T_c$ is obtained from the specific heat measurement and it is set to zero for those that exhibit no superconducting jump despite the fact that the resistivity goes to zero in some samples \cite{WrightJD12}. Therefore, the superconducting dome plot in Fig. \ref{fig1}(a) only includes the samples that show bulk superconductivity.

The raw data of specific heat are plotted in Fig. \ref{fig1}(b) for some of the samples. It is clear that the phonon contribution varies a lot for the different Co doping, which makes it impossible to use the specific heat of non-superconducting samples (e.g, x=0 or 0.2) as a reference to determine the electronic specific heat of superconducting samples as done in some other materials \cite{GofrykK10a,GofrykK10b,HardyF10}. 
To understand the specific heat of the non-superconducting samples, we consider 
 a Debye plus Einstein model assuming the total specific heat to be $ C = \gamma_NT + C_D+C_E$, 
 where $C_D = A_D(T/T_D)^3\int_0^{T_D/T}x^4e^x/(e^x-1)^2dx$ and $C_E = A_E(T_E/T)^2e^{T_E/T}/(e^{T_E/T}-1)^2$ are the specific heats from the Debye and Einstein models, respectively. 
 Fig. \ref{fig1}(c) shows the fitting results on the x=0.2 sample for the Debye+Einstein model and some other models. Fig. \ref{fig1}(d) further gives the differences between the raw data and the fitting results of various models, which unambiguously shows that the Debye+Einstein model gives the best fit to the data. We note that the parameters such as the Debye temperature and Einstein temperature in the fitting may not reflect the real phonon physics in this system. 
Since it is only possible to fit the normal-state and low-temperature data of the superconducting samples ( assuming that the superconducting gaps are fully opened), we also test the above method by removing the x = 0.2 data between 3 K and 20 K in the fitting and the result is consistent with that fitted with the whole temperature range. The value of $\gamma_N$ is manually adjusted for the superconducting samples to make sure that the entropy is conserved. In addition, the residual Sommerfeld coefficient $\gamma_0$ is obtained by fitting the low-temperature specific heat with $C = \gamma_0 T+\beta T^3$.

\begin{figure}[tbp]
\includegraphics[scale=.45]{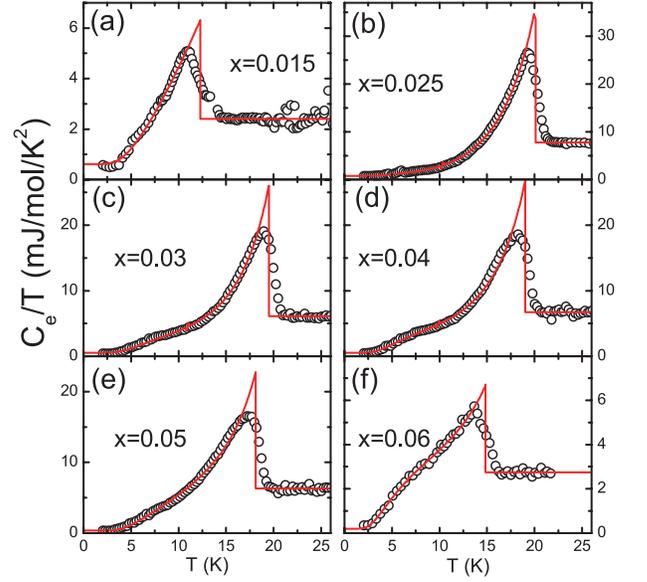}
\caption{(Color online) The electronic specific heat (black open circle) of serial samples obtained as described in the text. The red lines are the fitted results of one-gap BCS function except for the x=0 sample which shows no superconducting jump. The blue line in (h) is the subtracted data between 0 and 9 Tesla.
}
\label{fig2}
\end{figure}

\begin{figure}[tbp]
\includegraphics[scale=.45]{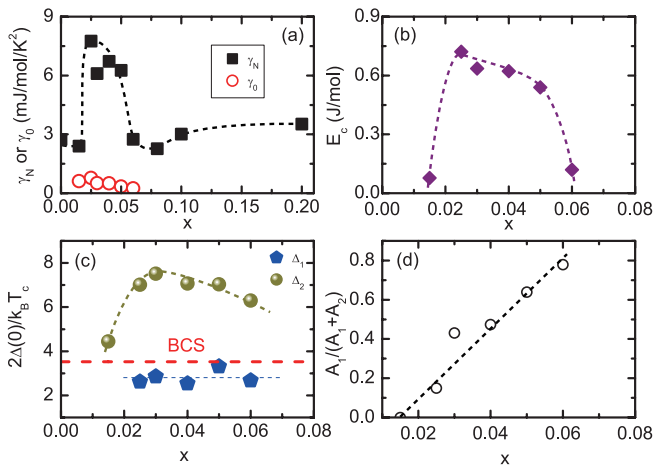}
\caption{(Color online) Doping dependence of (a) $\gamma_N$ ( black solid squares ) and $\gamma_0$ ( red open circles ), (b) the condensation energy, (c)  2$\Delta(0)/k_BT_c$ and (d) $A_1/(A_1+A_2)$. All the dashed lines are guided to the eye.
}
\label{fig3}
\end{figure}

The subtracted electronic specific heats of the superconducting samples are shown in Fig. \ref{fig2}. All the data except for the x = 0.015 can be well fitted by the two-gap BCS expression of the specific heat ( $C = A_1C_{BCS}(\Delta_1)+A_2C_{BCS}(\Delta_2)$) as shown by the solid lines \cite{GofrykK10a}. Only one gap is needed to fit the x = 0.015 data. It should be pointed out that we cannot rule out the existence of nodes or highly anisotropic gaps \cite{ChoK12,GeQ12} due to the limitation of our model. 
Fig. \ref{fig3}(a) shows the doping dependence of $\gamma_N$ and $\gamma_0$. Contrary to that in Ba(Fe$_{1-x}$Co$_x$)$_2$As$_2$ \cite{GofrykK10b}, $\gamma_0$ is much smaller than $\gamma_N$ for all the superconducting samples, suggesting that most of the electrons in NaFe$_{1-x}$Co$_x$As are condensed at 0 K. With increasing Co doping, $\gamma_N$ quickly increases and reaches its peak at the optimal doping with x = 0.025. Such suppression in the underdoped regime is most likely due to the opening of the SDW gap \cite{GofrykK10b}. Further increasing Co above 0.05 rapidly reduce $\gamma_N$ to a very low value for heavily over-doped samples ( $\gamma_N$ = 2.3 mJ/mol/K$^2$ for the x = 0.08 sample ). Surprisingly, the $\gamma_N$ goes back to more than 3 mJ/mol/K$^2$ for the x $\geq$ 0.1 samples.

The doping dependence of $\gamma_N$ may be strongly associated with the SDW gap and pseudogap as observed by the STM \cite{ZhouX12}. Since Co doping only shifts the Fermi level without significantly changing the band structures \cite{CuiST12}, we can quantitatively estimate the effect of the two gaps. Taking that N(0)(1+$\lambda$)=0.42$\gamma_N$/n \cite{StewartGR11} where n = 3 and N(0) $\approx$ 0.53 states/eV/atom \cite{DengS09}, we get that the coupling parameter $\lambda$ is about 0.9 for the x=0.025 sample. Such value is reasonable considering that no pseudogap is found near the optimal doping \cite{ZhouX12}. Supposing $\lambda$ does not change below 0.1 Co doping, we estimate that the DOS is suppressed about 60\% for both the x =0 and x = 0.06 samples, which is consistent with the STM results \cite{ZhouX12}. We note that the suppression of $\gamma_N$ in overdoped NaFe$_{1-x}$Co$_x$As is much larger than that in Ba(Fe$_{1-x}$Co$_x$)$_2$As$_2$ \cite{GofrykK10b}, which suggest that the latter may have a different origin. Increasing Co doping above 0.1 results in the depinning of the large "V"-shaped feature and thus the disappearance of the pseudogap \cite{ZhouX12}. Our measurements on the 0.1 and 0.2 samples show that $\lambda$ is close to zero assuming that there is no suppression of DOS at Fermi level, which accords with the fact that the system is close to a normal metal with weakly coupled electrons in this doping regime. 

Fig. \ref{fig3}(b) plots the doping dependence of the condensation energy $E_c$, which is obtained through $E_c = -\int^{T_c}_{0}\int^{T_c}_{0}C/TdTdT$. It is clear that $E_c$ is much smaller at either the x = 0.015 or x = 0.06 sample, consistent with the fact that the DOS at the Fermi level is strongly suppressed due to either the SDW gap or the pseudogap \cite{ZhouX12,CaiP12}.  

\begin{figure}[tbp]
\includegraphics[scale=.45]{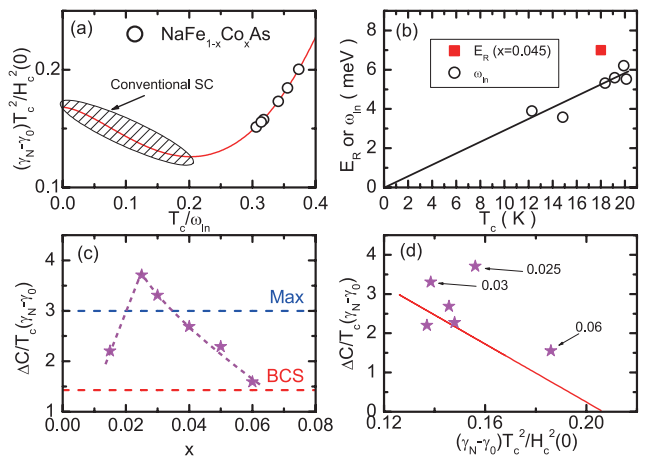}
\caption{(Color online) (a) Theoretical result of $(\gamma_N-\gamma_0)T_c^2/H_c^2(0)$ calculated from Eq. (2) as shown by the red line. The values of conventional superconductors fall into the shaded area. The open circles represent the values obtained in this paper which give the corresponding $T_c/\omega_{ln}$ (b) The $T_c$ dependence of the resonance energy $E_R$ and $\omega_{ln}$. The solid line is guided to the eye. (c) The doping dependence of $\Delta C/T_c(\gamma_N-\gamma_0)$. (d) The corresponding $(\gamma_N-\gamma_0)T_c^2/H_c^2(0)$ and $\Delta C/T_c(\gamma_N-\gamma_0)$ for each sample. The solid line is the linear relationship between these two values as calculated from Eq. (1) and (2).
}
\label{fig4}
\end{figure}

The doping dependences of the two superconducting gaps and the relevant ratio of the small gap are shown in Fig. \ref{fig3}(c) and \ref{fig3}(d) respectively. The values of the larger gap above x = 0.025 are more or less consistent with the results of ARPES and STM experiments where only one gap is observed \cite{LiuZH11,ZhouX12,YangH12}. The existence of the smaller gap and its increasing contribution to the electronic specific heat are missing in those experiments. Since the tunneling matrix element of the M-centered bands may be strongly suppressed for a good surface in the STM experiment \cite{HoffmanJE11}, the small gap may exist around the M point with a non-zero $k_z$ value \cite{LiuZH11}. For the x = 0.015 sample where the AF order presents, it is not clear why a much smaller gap is obtained from the specific heat data \cite{CaiP12,GeQ12}.

The two dimensionless ratios in Eq. (1) and (2) can be derived from the above experimental data. To eliminate the effect of the residual electronic specific heat, we replace $\gamma_N$ in the ratios to $\gamma_N-\gamma_0$. Fig. 4(a) shows the doping dependence of $(\gamma_N-\gamma_0)T_c^2/H_c^2(0)$ by taking $H_c(0)^2 = 8\pi E_c$. For the conventional superconductors, the ratios of many materials are within the shaded area centering the red line in Fig. \ref{fig4}(a) calculated by Eq. (1) \cite{CarbotteJP90}. By assuming that it is on the other side of the curve in our case (a value of $T_c/\omega_{ln}$ much smaller than 0.2 will result in a bosonic energy that has not been observed in other experiments), we are able to obtain $T_c/\omega_{ln}$ ( the open circles in Fig. \ref{fig4}(a) ). Fig. 4(b) shows the $T_c$ dependence of $\omega_{ln}$, which gives $\omega_{ln} = 3.38k_BT_c$. Interestingly, the resonance energy in the NaFe$_{0.955}$Co$_{0.045}$As is very close to the value of $\omega_{ln}$ \cite{ZhangC12}, suggesting the resonance mode may play as bosons in the superconductivity of NaFe$_{1-x}$Co$_{x}$As. It will be interesting to compare the resonance energy in heavily underdoped and overdoped samples with the $\omega_{ln}$ obtained here. The large values of $T_c/\omega_{ln}$ suggest the strong-coupling nature of NaFe$_{1-x}$Co$_{x}$As noting that the ratio only extends up to about 0.25 in the conventional superconductors. We note that the above $\omega_{ln}$ can also give a value of $2\Delta(0)/k_BT_c$ larger than 5 as suggested by the equation (4.1) in Ref. \cite{CarbotteJP90}. 

The doping dependence of the normalized specific heat jump $\Delta C/T_c(\gamma_N-\gamma_0)$ is shown in Fig. \ref{fig4}(c). In the case where the two-gap BCS function cannot give a good fit near $T_c$, a sing-gap BCS function is used to just fit the data near $T_c$ to obtain an accurate specific heat jump. A dome-like feature is seen as that in Ba(Fe$_{1-x}$Co$_x$)$_2$As$_2$ \cite{GofrykK10b}. Surprisingly, the largest value of $\Delta C/T_c(\gamma_N-\gamma_0)$ is 3.7 for x = 0.025, which is much larger than those found in other systems \cite{PopovichP10,NojiT12}. While the values of $(\gamma_N-\gamma_0)T_c^2/H_c^2(0)$ seem to be reasonable, Eq. (2) fails to calculate $\Delta C/T_c(\gamma_N-\gamma_0)$ at the optimal doping. This is clearer by plotting these two ratios together as shown in Fig. \ref{fig4}(d). The data falling on the solid line suggest that they can be calculated from each other according to Eq. (1) and (2). It is clear that strong deviation occurs near the optimal doping. Since $(\gamma_N-\gamma_0)T_c^2/H_c^2(0)$ is close to each other except for that of x = 0.06, such deviation is not due to the insufficiency in calculating the two ratios by the perturbation method \cite{CarbotteJP90}. For the x = 0.06 sample, it is possible that a more accurate method may give a better result or the pseudogap-like phase may has something to do with the mismatch.

A large specific heat jump in the Eliashberg model is a result that the superconducting gap opens up more rapidly just below $T_c$ than it does in the BCS theory \cite{CarbotteJP90}, which will only give a maximum value of about 3 as seen in Eq. (2). In the case of NaFe$_{1-x}$Co$_{x}$As, one may has to consider a very weak temperature dependence of the gap \cite{LiuZH11}. On the other hand, the ratio $\gamma_NT_c^2/H_c^2(0)$ is associated with the condensation of the Cooper pairs at zero K, which may not contradict with what happens near $T_c$. After all, a very important assumption in the strong-coupling theory is that the boson spectrum is fixed while the spin fluctuations in the Fe-based superconductors strongly evolve with changing temperature.

It is also found in the heavy-fermion materials CeCoIn$_5$ \cite{PetrovicC01} and CeIrSi$_3$ \cite{TateiwaN08} that $\Delta C/T_c\gamma_N$ is lager than 4. 
This is consistent with theories associated with the strong localized spin fluctuations \cite{IkedaH03,BangY04,KhodelVA05}, indicating that 
spin fluctuations may result in the largest enhancement of the specific heat jump near optimal doping \cite{VavilovMG11}.  

In conclusion, we test the validity of the Eliashberg formalism in the NaFe$_{1-x}$Co$_x$As system by deriving $\gamma_NT_c^2/H_c^2(0)$ and $\Delta C/T_c(\gamma_N-\gamma_0)$ from the specific heat measurements. Our results show that while the former value is nicely associated with the neutron spin resonance through Eq. (1), the latter value is beyond the Eliashberg theory near optimal doping. Therefore, the pairing mechanism in NaFe$_{1-x}$Co$_x$As may be understood within the boson-exchange mechanism but the disappearance of the superconductivity near the optimal doping should be considered with significant modification of the theory. 

The authors would like to thank Yifeng Yang, Tao Xiang, Yayu Wang, Haihu Wen, Daoxin Yao and Lei Shan for helpful discussions.
This work is supported by Chinese Academy of Science, 973 Program (2010CB833102, 2010CB923002, 2012CB821400, 2011CB921701). The single crystal
growth at University of Tennessee was supported by U.S. DOE
BES under Grant No. DE-FG02-05ER46202 (P.D.).

\bibliography{NaCoFeAs}

\end{document}